\documentclass[12pt,preprint2]{emulateapj}

\usepackage{lineno}
\usepackage{natbib}
\usepackage{amsmath}
\usepackage{graphicx}
\usepackage{wasysym}
\usepackage{txfonts}
\usepackage{xspace}
\usepackage{url}
\usepackage{textcomp}
\usepackage{multirow}
\usepackage{lipsum}

\newcommand{\gx}{GX~339$-$4\xspace}
\newcommand{\grs}{GRS~1739$-$278\xspace}

\newcommand{\swift}{\textsl{Swift}\xspace}

\newcommand{\xmm}{\textsl{XMM-Newton}\xspace}

\newcommand{\nustar}{\textsl{NuSTAR}\xspace}

\newcommand{\asec}{\ensuremath{''}\xspace}

\newcommand{\snr}{S/N\xspace}

\newcommand{\msun}{\ensuremath{\text{M}_{\odot}}\xspace}

\newcommand{\redchi}{\ensuremath{\chi^{2}_\text{red}}\xspace}

\newcommand{\feka}{\ensuremath{\mathrm{Fe}~\mathrm{K}\alpha}\xspace}

\newcommand{\ledd}{\ensuremath{{L}_{\mathrm{Edd}}}\xspace}
\renewcommand{\deg}{\ensuremath{^\circ}}

\newcommand{\rin}{\ensuremath{R_\text{in}}\xspace }
\newcommand{\rg}{\ensuremath{R_\text{g}}\xspace }
\newcommand{\risco}{\ensuremath{R_\text{ISCO}}\xspace }
\newcommand{\rrefl}{\ensuremath{\mathcal{R}_\text{refl}}\xspace}

\shorttitle{\grs observed with \xmm and \nustar}
\shortauthors{F\"urst et al.}

\begin{document}

\title{\grs observed at very low luminosity with \xmm and \nustar}

\author{F.~F\"urst\altaffilmark{1}}
\author{J.~A.~Tomsick\altaffilmark{2}}
\author{K.~Yamaoka\altaffilmark{3,4}}
\author{T.~Dauser\altaffilmark{5}}
\author{J.~M.~Miller\altaffilmark{6}}
\author{M.~Clavel\altaffilmark{2}}
\author{S.~Corbel\altaffilmark{7,8}}
\author{A.~Fabian\altaffilmark{9}}
\author{J.~Garc\'ia\altaffilmark{10}}
\author{F.~A.~Harrison\altaffilmark{1}}
\author{A.~Loh\altaffilmark{7}}
\author{P.~Kaaret\altaffilmark{11}}
\author{E.~Kalemci\altaffilmark{12}}
\author{S.~Migliari\altaffilmark{13,14}}
\author{J.~C.~A.~Miller-Jones\altaffilmark{15}}
\author{K.~Pottschmidt\altaffilmark{16,17}}
\author{F.~Rahoui\altaffilmark{18,19}}
\author{J.~Rodriguez\altaffilmark{7}}
\author{D.~Stern\altaffilmark{20}}
\author{M.~Stuhlinger\altaffilmark{13}}
\author{D.~J.~Walton\altaffilmark{20,1}}
\author{J.~Wilms\altaffilmark{5}}

\altaffiltext{1}{Cahill Center for Astronomy and Astrophysics, California Institute of Technology, Pasadena, CA 91125, USA}
\altaffiltext{2}{Space Sciences Laboratory, University of California, Berkeley, CA 94720, USA}
\altaffiltext{3}{Solar-Terrestrial Environment Laboratory, Nagoya University, Furo-cho, Chikuka-ku, Nagoya, Aichi
464-8601, Japan}
\altaffiltext{4}{Division of Particle and Astrophysical Science, Graduate School of Science, Nagoya University,
Furo-cho, Chikuka-ku, Nagoya, Aichi 464-8602, Japan}
\altaffiltext{5}{Dr. Karl-Remeis-Sternwarte and ECAP, Sternwartstr. 7, 96049 Bamberg, Germany}
\altaffiltext{6}{Department of Astronomy, The University of Michigan, Ann Arbor, MI 48109, USA}
\altaffiltext{7}{Laboratoire AIM (CEA/IRFU-CNRS/INSU-Universit\'e Paris Diderot), CEA DSM/IRFU/SAp, 91191 Gif-sur-Yvette, France}
\altaffiltext{8}{Station de Radioastronomie de Nan\c{c}ay, Observatoire de Paris, PSL Research University, CNRS, Univ. Orl\'{e}ans, 18330 Nan\c{c}ay, France}
\altaffiltext{9}{Institute of Astronomy, Madingley Road, Cambridge CB3 0HA, UK}
\altaffiltext{10}{Harvard-Smithsonian Center for Astrophysics, Cambridge, MA 02138, USA}
\altaffiltext{11}{Department of Physics and Astronomy, University of Iowa, Iowa City, IA 52242, USA}
\altaffiltext{12}{Faculty of Engineering and Natural Sciences, Sabanc{\i} University, Orhanl\i-Tuzla, 34956 Istanbul, Turkey}
\altaffiltext{13}{European Space Astronomy Centre (ESAC), 28692 Villanueva de la Ca\~nada, Madrid, Spain}
\altaffiltext{14}{Department of Quantum Physics and Astrophysics \& Institute of Cosmos Sciences, University of Barcelona,  08028 Barcelona, Spain}
\altaffiltext{15}{International Centre for Radio Astronomy Research - Curtin University, GPO Box U1987, Perth, WA 6845, Australia}
\altaffiltext{16}{CRESST, Department of Physics, and Center for Space Science and
Technology, UMBC, Baltimore, MD 21250, USA}
\altaffiltext{17}{NASA Goddard Space Flight Center, Greenbelt, MD 20771, USA}
\altaffiltext{18}{European Southern Observatory, 85748 Garching bei Munchen, Germany}
\altaffiltext{19}{Department of Astronomy, Harvard University, Cambridge, MA 02138, USA}
\altaffiltext{20}{Jet Propulsion Laboratory, California Institute of Technology, Pasadena, CA 91109, USA}

\begin{abstract}

We present a detailed spectral analysis of  \xmm and \nustar observations of the accreting transient black hole \grs during a very faint low hard state at $\sim$0.02\% of the Eddington luminosity (for a distance of 8.5\,kpc and a mass of 10\,\msun). The broad-band X-ray spectrum between 0.5--60\,keV can be well-described by a power law continuum with an exponential cutoff. The continuum is unusually hard for such a low luminosity, with a  photon index of $\Gamma=1.39\pm0.04$. We find  evidence for an additional reflection component from an optically thick accretion disk at the 98\% likelihood level. The reflection fraction is  low with $\rrefl=0.043^{+0.033}_{-0.023}$. In combination with  measurements of the spin and inclination parameters made with \nustar during a brighter hard state  by Miller and co-workers, we seek to constrain the accretion disk geometry. Depending on the assumed emissivity profile of the accretion disk, we find a truncation radius of 15--35\,$\rg$ (5--12\,$\risco$) at the 90\% confidence limit. 
These values depend strongly on the assumptions and we discuss possible systematic uncertainties.

\end{abstract}


\keywords{stars: black holes --- X-rays: binaries --- X-rays: individual (GRS 1739-278) --- accretion, accretion disks}

\section{Introduction}
\label{sec:intro}
Galactic black hole (BH) transients typically undergo a very characteristic pattern during an outburst: during the first part of the rise, up to luminosities around 10\% of the Eddington luminosity (\ledd), they are in a so-called low/hard state. In this state the X-ray spectrum is dominated by a power law with a photon index $\Gamma$ between $\approx$1.4--1.8 with almost no contribution from the thermal accretion disk spectrum. At higher Eddington rates the source switches to the high/soft state, where a  steeper power law is observed  and the thermal accretion disk dominates the soft X-ray spectrum \citep[see, e.g.,][for a description of BH states]{remillard06a}.
Compelling evidence exists that in the soft state the accretion disk  extends to the innermost stable circular orbit (ISCO), enabling spin measurements through relativistically smeared reflection features and thermal continuum measurements \citep[e.g.,][]{nowak02a, miller02a, steiner10a, mcclintock14a, petrucci14a,  kolehmainen14a, miller15a, parker16a}.

At the end of an outburst the source transitions back to the low/hard state, albeit typically at much lower luminosities ($\approx$1--4\%\,\ledd) in a hysteretic behavior \citep[see, e.g.,][]{maccarone03a,kalemci13a}. 
It has been postulated that the accretion disk recedes, i.e., the inner accretion disk radius \rin is no longer at the ISCO. Instead the inner regions are replaced by an advection dominated accretion flow (ADAF) in the inner few gravitational radii \citep[e.g.,][]{narayan95a, esin97a}. 
Many observational results in a sample of different sources
are at least qualitatively consistent with such a truncated disk as measured by, e.g., the frequency and width of quasi-periodic oscillations or multi-wavelength spectroscopy \citep[see, e.g.,][]{zdziarski99a, esin01a,kalemci04a,tomsick04a}.

It is still not clear, however, at what luminosity the truncation occurs and how it is triggered.  There have been several reports of broad iron 
lines (implying a non-truncated disk) in the brighter part of the low/hard state ($>1$\%\,\ledd) for 
\gx \citep{miller06a,reis11a,allured13a} as well as for other systems 
\citep{reis10a,reynolds10a}, including \grs \citep[hereafter M15]{miller15a}.

Studies conducted recently mostly claim evidence for moderate (tens of gravitational radii $\rg$) 
truncation at intermediate luminosities ($\approx$0.5--10\%\,\ledd) in the low/hard state
\citep{shidatsu11a,allured13a,petrucci14a,plant14a}.  
At a luminosity of $L=0.14\%\,\ledd$ in \gx, \citet{tomsick09a} measured a narrow \feka line, indicating a significant truncation.
While this suggests
that gradual truncation may occur, it is not clear that $\rin$ is
only set by the luminosity  \citep{petrucci14a, kolehmainen14a, garcia15a}. A more complex situation than a simple correlation with luminosity is also supported by recent measurements of the disk truncation at $\sim10\,\rg$ in \gx during intermediate states, i.e., during state transitions,  at luminosities of 5--10\%\,\ledd \citep{tamura12a, gx339IHS}.

 Besides the truncation radius, the geometry of the hot electron gas, or corona, is still unclear. It is very likely compact, and it has been postulated that it might be connected to the base of the jet, though a commonly accepted model has not yet emerged \citep[see, e.g.][]{markoff05a, reis13a}.  \nustar and \swift observations of \gx in the low/hard state found that the reflector seems to see a different continuum than the observer, i.e., a hotter part of the corona \citep{gx339}. This indicates a temperature gradient and a complex structure of the corona and seems to be independent of the spectral state \citep{parker16a}. 

It is clear from previous studies that the largest truncation radius is expected at the lowest luminosities, i.e., at the end and beginning of an outburst. High quality data in this state are traditionally difficult to obtain, given the low flux and necessary precise scheduling of the observations before the source vanishes into quiescence. With a combination of \xmm \citep{xmmref} and the \textsl{Nuclear Spectroscopy Telescope Array}  \citep[\nustar,][]{harrison13a}, however, such observations are now possible.
 
\begin{figure}
\includegraphics[width=0.95\columnwidth]{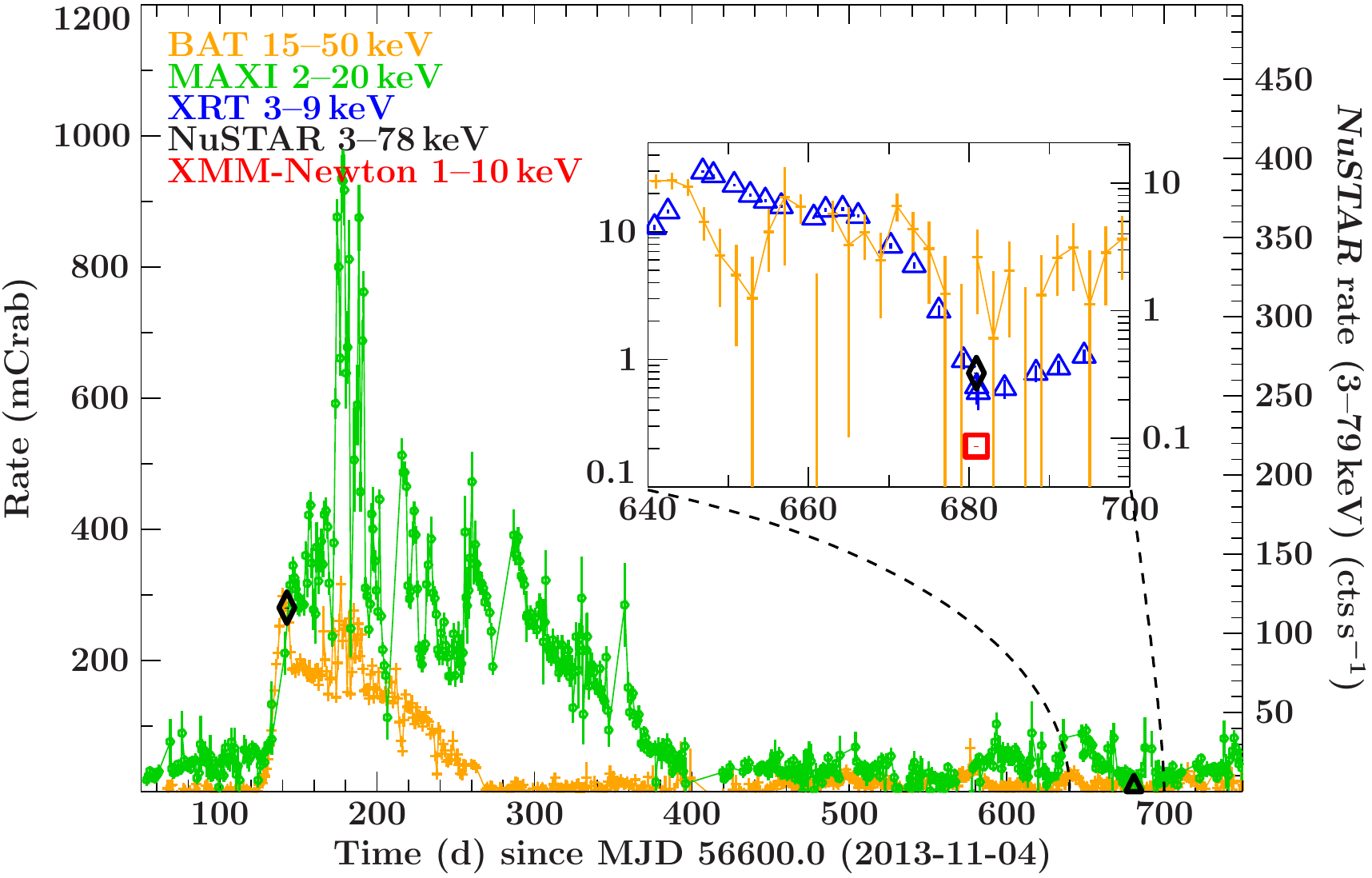}
\caption{\swift/BAT \citep[15--50\,keV, orange;][]{swiftbatref} and MAXI/GSC \citep[2--20\,keV, green; ][]{maxiref} monitoring light curve of \grs. The \nustar observations (3--79\,keV) are marked by black diamonds, the one presented by \citet{miller15a} occurred around 150\,d, the one presented here around 680\,d. All data are shown in observed (i.e., absorbed) count-rates rescaled to mCrab fluxes in the respective energy band of the instrument. 
The right-hand $y$-axis gives the average  measured \nustar count-rate of the observation. The inset shows a zoom-in on the 2015 data, including \swift/XRT \citep{swiftxrtref} data (3--9\,keV, blue triangles) and the \xmm observation (1--10\,keV, red square). 
Due to the crowded source region the MAXI data suffer from increased background of about 40\,mCrab and are therefore not shown in the inset.
Note that the inset $y$-axes are scaled logarithmically.
}
\label{fig:batlc}
\end{figure}

Here we report on \xmm and \nustar observations of the BH transient \grs in the declining phase of its very long outburst in 2014/2015 (Figure~\ref{fig:batlc}). 
\grs is a transient BH candidate, discovered by \textsl{Granat} \citep{paul96a, vargas97a}. It is most likely located close to the Galactic Center at a distance of $\approx8.5$\,kpc. The large extinction \citep[$A_V=14\pm2$,][]{greiner96a} makes a spectral identification of the companion difficult, but from photometric data, \citet{marti97a} and \citet{chaty02a}  infer a late-type main-sequence star of at least F5~V or later.

\grs was classified as a BH candidate given its similarity in spectral evolution to other transient BHs as well as the presence of a very strong 5\,Hz QPO in the soft-intermediate state \citep{borozdin98a,borozdin00a}.

 During the beginning of the 2014/2015 outburst, \nustar measured a strong reflection spectrum and a relativistically broadened iron line in a bright low/hard state (M15). These authors could constrain the size of the corona, assuming a lamppost model, to be $<22\,\rg$ and the truncation radius to $\rin = 5^{+3}_{-4}\,\rg$. 
In the lamppost geometry  the corona is assumed to be a point-like source located on the spin axis of the BH and shining down onto the accretion disk \citep{matt91a, dauser13a}.
 The luminosity during this observation was around 8\%\ledd (assuming a canonical mass of 10\,\msun), at which no truncation of the accretion disk is expected.

After the first \nustar observation, the source continued with a typical outburst evolution and faded to very low luminosities around MJD~57000. However, it probably never reached quiescent levels and \swift/XRT and BAT monitoring indicated that it  also did not switch back to a stable low/hard state.
A detailed description of the evolution will be presented by  Loh et al. (in prep.).
Around MJD~57272 the monitoring data indicated a stable transition to the low/hard state had occurred, confirmed by a brightening in the radio.  We then triggered simultaneous \xmm and \nustar observations  to observe a very faint hard state, and found \grs at $\sim$0.02\%\,\ledd.

The rest of the letter is structured as follows: in Section~\ref{sec:data} we  describe the data reduction and calibration. In Section~\ref{sec:spec} we present the spectral analysis and compare it to results by M15. In the last section, Section~\ref{sec:disc}, we discuss our results and put them into context.
 
\section{Data reduction and observation}
\label{sec:data}
\subsection{\nustar}
\label{susec:nustar}
\nustar observed \grs on MJD~57281 (ObsID 80101050002) for a good exposure time, after standard screening, of 43\,ks per module. We extracted the \nustar data using HEASOFT v6.15 and the standard \texttt{nupipeline} v1.4.1 from a  50\asec region centered on the J2000 coordinates of \grs. On both focal plane modules (FPMs) the source was located in an area of enhanced background due to stray-light from sources outside the field-of-view, dominated by GX~3+1. We tested different background regions and found that the exact choice only marginally influences the source spectrum. We obtained good agreement between FPMA and FPMB. Despite the high background level we obtained a detection up to 60\,keV. We used \nustar data between 3--60\,keV and rebinned them to a  signal-to-noise ratio (\snr) of 6 per bin and at least 2 channels per bin (Figure~\ref{fig:spec}\textit{a}).

\subsection{\xmm}
\label{susec:xmm}
We obtained simultaneous \xmm observations with a good exposure time of 79\,ks in EPIC-pn \citep{pnref}, using the  timing mode (ObsID 0762210201). \xmm data were extracted using SAS v14.0.0. The source spectrum was extracted from columns RAWX 33--42 and the background from columns RAWX 50--60 using only single and double events (PATTERN 0--4). The first 15\,ks of the observation were strongly contaminated by background flares, and we excluded these data. The background continued to be elevated throughout the whole observation, in particular influencing the spectrum below 1\,keV. In the remainder of the paper, we therefore use EPIC-pn data between 0.6--10\,keV, rebinned to a \snr  of 5 with at least 5 channels per bin.

We also obtained EPIC-MOS \citep{mosref} data in timing mode. Due to a hot column, calibration of the MOS\,1 timing mode is difficult and we therefore ignore these data. For the MOS\,2 data, the source spectrum was extracted from columns RAWX 294--314  and the background from columns RAWX 260--275 using only single events (PATTERN=0) with FLAG=0. 
MOS\,2 data add up to a good exposure time of 35\,ks and were rebinned to a \snr of 5 with at least 3 channels per bin between 0.7--10\,keV. They agree very well with the EPIC-pn data (Figure~\ref{fig:spec}).

\section{Spectral Analysis}
\label{sec:spec}
Using the  Interactive Spectral Interpretation System \citep[ISIS v1.6.2,][]{houck00a} we fit the \xmm and \nustar spectra simultaneously. Uncertainties are reported at the 90\% confidence level unless otherwise noted.  We allowed for a cross-calibration constant  ($CC$) between the instruments to take differences in absolute flux calibration into account. All fluxes are given with respect to \nustar/FPMA. The other instruments are within a few percent of these values, besides MOS\,2, which measures fluxes up to 15\% lower. This discrepancy is  within the expected uncertainty of the MOS timing mode.

We model the absorption using an updated version of the \texttt{tbabs}\footnote{\url{http://pulsar.sternwarte.uni-erlangen.de/wilms/research/tbabs/}} model and its corresponding abundance vector  as described by \citet{wilms00a} and cross-sections by \citet{verner96a}. As found by M15 and other previous works, the column density is around $2\times10^{22}$\,cm$^{-2}$,  in agreement with the estimates from the dust scattering halo found around \grs \citep{greiner96a}.

\begin{figure}
\includegraphics[width=0.95\columnwidth]{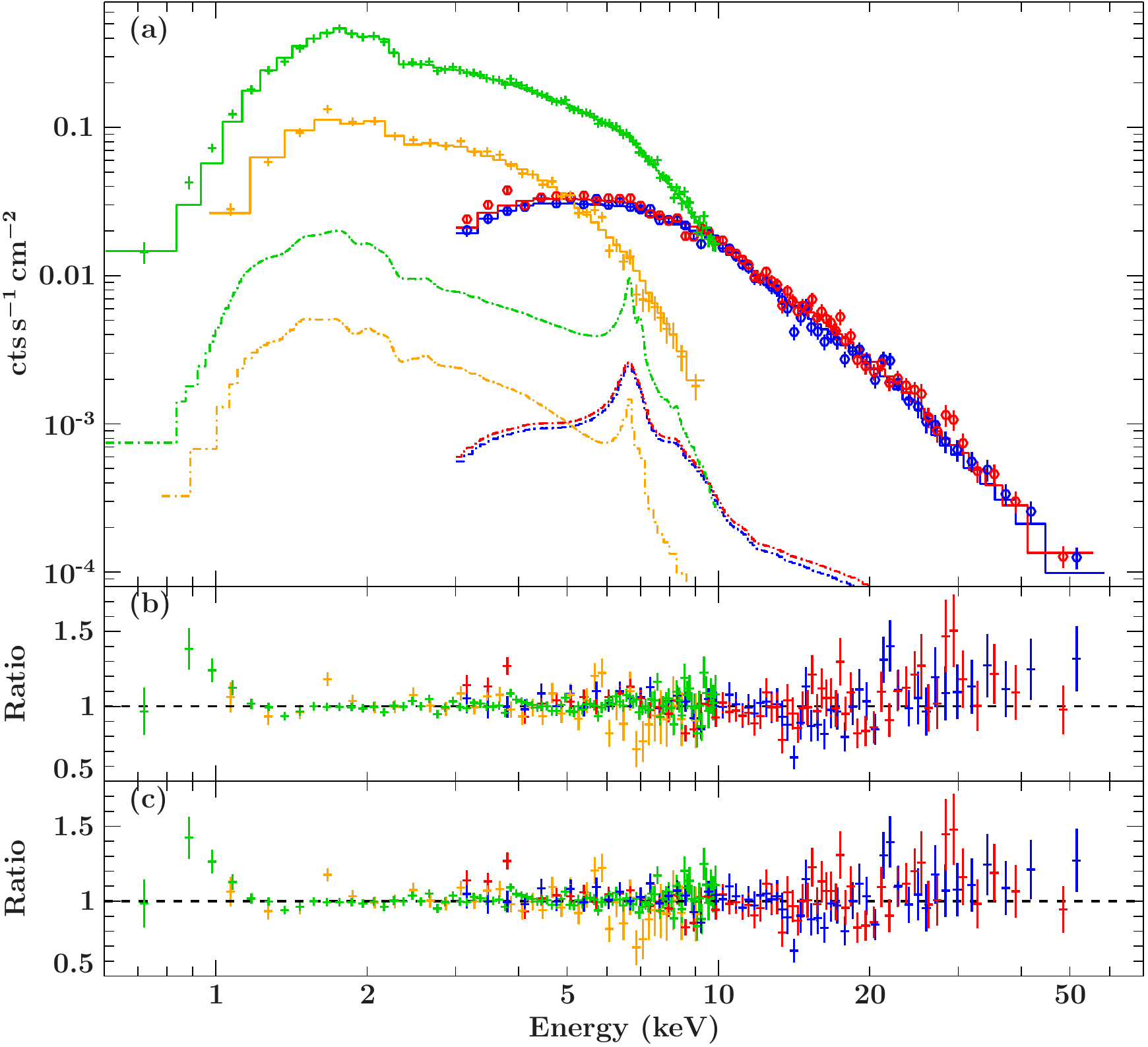}
\caption{\textit{(a)} Data and best-fit \texttt{xillver} model. \xmm/EPIC-pn is shown in green, MOS\,2 in orange, \nustar/FPMA in red and FPMB in blue. The dashed lines show the contribution of the reflection in each instrument. \textit{(b)}  Residuals to the cutoff-power law model. \textit{(c)} Residuals to the reflection (\texttt{xillver}) model. Data were rebinned for visual clarity.
}
\label{fig:spec}
\end{figure}

Using an absorbed power law continuum with an exponential cutoff  provides a statistically acceptable fit, with $\redchi=1.08$ ($\chi^2=1023$) for 946 degrees of freedom (d.o.f.). The best-fit values are given in Table~\ref{tab:bestfit15} and the residuals are shown in Figure~\ref{fig:spec}b. Small deviations around 1\,keV can be attributed to known calibration uncertainties in the EPIC instruments.

Compared to the earlier observation discussed by M15, the spectrum of the later observation discussed here is significantly harder, with a lower photon index $\Gamma$ and a higher folding energy $E_\text{fold}$ (labeled $E_\text{cut}$ in the \texttt{cutoffpl} model and in  M15). This is not only true when compared to the simple cutoff power law model of M15, which does not provide an adequate fit to their data, but also when compared to the underlying continuum when adding an additional reflection component  (see Table~1 in M15).

The \texttt{cutoffpl} is continuously curving (even far below the folding energy) and does not necessarily accurately describe a Comptonization spectrum \citep{cena, fabian15a}. We therefore also tested the Comptonization model \texttt{nthcomp} \citep{zdziarski96a,zycki99a} and find a comparable fit with $\redchi=1.08$ ($\chi^2 = 1022$) for 945 d.o.f. (Table~\ref{tab:bestfit15}). We find a plasma temperature of $kT_\text{e} = 15.5^{+6.3}_{-2.7}$\,keV, which, when multiplied with the expected factor of 3, agrees well with the measured folding energy of the \texttt{cutoffpl}.

We next search for signatures of reflection, which is present in all low/hard state spectra of accreting black holes, even at low luminosities \citep[see, e.g.,][]{tomsick09a,gx339}. To model the reflection we use the \texttt{xillver} model v0.4a \citep{garcia10a, garcia13a}, which self-consistently describes the iron line and Compton hump. The model is based on a cutoff power law as the input continuum and we therefore also use the cutoff power law to describe the continuum spectrum. 

With this model we find a statistically  good fit with $\redchi=1.06~(\chi^2=1005)$ for 943 d.o.f. We show this model with the data and the contribution of the reflection  in Figure~\ref{fig:spec}\textit{a} and its residuals in Figure~\ref{fig:spec}\textit{c}.
This is an improvement of  $\Delta\chi^2 = 15$ for 3 fewer degrees of freedom. According to the sample-corrected Akaike Information Criterion \citep[AIC,][]{akaike74a},  this is a significant improvement of $\Delta$AIC=8.8, i.e., at $>$98\% likelihood \citep{burnham11a}.

We find a low, but well constrained reflection fraction of $\rrefl=0.045^{+0.044}_{-0.022}$ and a  high ionization parameter of $\log(\xi/(\text{erg\,cm\,s}^{-1}))  = 3.22^{+0.43}_{-0.27}$. The iron abundance $A_\text{Fe}$ is not well constrained but seems to prefer values $>2.5$ solar, relative to the solar abundances by \citet{grevesse98a}, on which the \texttt{xillver} model is based (Table~\ref{tab:bestfit15}). Fixing the iron abundance to 1 times solar results in a slightly worse fit with $\redchi=1.07~(\chi^2=1013$) for 944 d.o.f., but none of the other parameters changes significantly. We cannot constrain the ratio between neutral and ionized iron due to small contribution of the reflection component to the overall spectrum.

\begin{deluxetable*}{rlllll}
\tablewidth{0pc}
\tablecaption{Best-fit model parameters.\label{tab:bestfit15}}
\tablehead{\colhead{Parameter}  & \colhead{Cutoffpl} & \colhead{Nthcomp} & \colhead{Xillver} & \colhead{Relxill} & \colhead{Relxilllp}}
\startdata
 $ N_\text{H}~(10^{22}\,\text{cm}^{-2})$ & $2.13\pm0.05$ & $1.44\pm0.06$ & $2.17^{+0.07}_{-0.05}$ & $2.16^{+0.06}_{-0.05}$ & $2.16^{+0.06}_{-0.05}$ \\
 $ \mathcal{F}~(10^{-11}\,\text{erg}\,\text{cm}^{-2}\,\text{s}^{-1})\tablenotemark{a}$ & $2.89\pm0.06$ & $2.79\pm0.05$ & $2.90^{+0.07}_{-0.04}$ & $2.91\pm0.06$ & $2.91\pm0.06$ \\
 $ \Gamma$ & $1.40\pm0.04$ & $1.637^{+0.016}_{-0.014}$ & $1.409^{+0.038}_{-0.026}$ & $1.404^{+0.030}_{-0.031}$ & $1.404^{+0.030}_{-0.031}$ \\
 $ E_\text{fold}/kT~(\text{keV})$ & $56^{+15}_{-10}$ & $15.5^{+6.3}_{-2.7}$ & $61^{+20}_{-10}$ & $58^{+14}_{-10}$ & $58^{+14}_{-10}$ \\
 $ A_\text{Fe}$ & --- & --- & $5.0^{+5.1}_{-2.5}$ &1.5\tablenotemark{c} & 1.5\tablenotemark{c}  \\
 $ \log\xi~(\text{erg\,cm\,s}^{-1})$ & --- & --- & $3.22^{+0.43}_{-0.27}$ & $3.22^{+0.23}_{-0.46}$ & $3.24^{+0.22}_{-0.49}$ \\
 $ \mathcal{R}_\text{refl}$ & --- & --- & $0.045^{+0.044}_{-0.022}$ & $0.08^{+0.06}_{-0.05}$ & $0.099$\tablenotemark{b}  \\
  $ i $ & --- & --- & $32.5\deg\tablenotemark{c}$ & $32.5\deg$\tablenotemark{c} &$32.5\deg$\tablenotemark{c} \\
 $ R_\text{in}~(R_\text{g})$ & --- & --- & --- & $>15$ & $>35$ \\
 $ R_\text{out}~(R_\text{g})$  & --- & --- & --- & $400$\tablenotemark{c} & $400$\tablenotemark{c} \\
 $ H~(R_\text{g})$ & --- & --- & --- & --- & $30_{-27}^{+100}$ \\
  $ q $ & --- & --- & --- & 3\tablenotemark{c} & --- \\
 $ a$ & --- & --- &--- & $0.8$\tablenotemark{c} & $0.8$\tablenotemark{c} \\
  $ CC_B$ & $0.979^{+0.025}_{-0.024}$ & $0.979^{+0.025}_{-0.024}$ & $0.984^{+0.022}_{-0.029}$ & $0.980^{+0.025}_{-0.024}$ & $0.980^{+0.025}_{-0.024}$ \\
 $ CC_\text{pn}$ & $0.960^{+0.020}_{-0.019}$ & $0.949\pm0.020$ & $0.958^{+0.015}_{-0.026}$ & $0.954\pm0.020$ & $0.954\pm0.020$ \\
 $ CC_\text{MOS}$ & $0.884^{+0.024}_{-0.023}$ & $0.872^{+0.024}_{-0.023}$ & $0.884^{+0.020}_{-0.029}$ & $0.880^{+0.024}_{-0.023}$ & $0.880^{+0.024}_{-0.023}$ \\
\hline$\chi^2/\text{d.o.f.}$   & 1022.70/946& 1022.48/945& 1005.78/943& 1012.06/943& 1012.31/943\\
$\chi^2_\text{red}$   & 1.081& 1.082& 1.067& 1.073& 1.073\enddata
\tablenotetext{a}{between 1--30\,keV}
\end{deluxetable*}

While the phenomenological models presented above provide a statistically very good fit, they do not contain information about the geometry of the X-ray producing region. To obtain information about the geometry we need to study the strong relativistic effects close to the BH, in particular the relativistic broadening of the reflection features. These features have been used by M15 in the bright hard state data to measure the spin of the BH in \grs to be $a=0.8\pm0.2$ and constrain the radius of the inner accretion disk to be close to the ISCO.

Due to the low count rates and low reflection strength, our data do not allow us to constrain all parameters of the relativistic smearing models. We therefore fix values that are unlikely to change on time-scales of the outburst, namely the inclination $i$ and the iron abundance $A_\text{Fe}$, to the values found by M15 for the \texttt{relxilllp} model: $i=32.5^\circ$ and $A_\text{Fe}=1.5$. We fix the spin to  $a=0.8$, the best-fit value of the \texttt{relxill} model by M15, as it was unconstrained in their lamppost geometry (\texttt{relxilllp}) model. By fixing the inclination, we ignore possible effects of a warped disk.

We model the relativistic effects using the \texttt{relxill} model \citep{dauser13a, garcia14a} with the emissivity described by a power law with an index of $3$, which is appropriate for a standard Shakura-Sunyaev accretion disk and  an extended corona \citep{dabrowski97a}. We also set the outer disk radius to $r_\text{out} = 400\,\rg$. This model gives a good fit with $\redchi=1.07~(\chi^2=1012)$ for 943 d.o.f., and its  best-fit parameters  are shown in Table~\ref{tab:bestfit15}. 
This fit is statistically slightly worse compared to the \texttt{xillver} model, but presents the more physically realistic description of the spectrum.
The main driver of reduced statistical quality is the iron abundance, which we held fixed. If we allow it to vary, we find a fit with $\redchi=1.07~(\chi^2=1332)$ for 1246 d.o.f., i.e.,  the same as for the \texttt{xillver} model. However, as in the \texttt{xillver} model, the iron abundance is only weakly constrained and the other parameters do not change significantly. Thus, we keep it fixed at the better constrained value from M15 for the remainder of this work.
 We only obtain a lower limit on the inner accretion disk radius, $\rin> 15\,\rg$. 

Allowing for a variable emissivity index does not improve the fit significantly and results in a similar constraint for the inner radius ($\rin > 15\,\rg$). The emissivity index itself is not constrained between $3\leq q \leq 10$. The often used broken power law emissivity profile can therefore not be constrained either, in particular because the expected break radius is smaller than the inner accretion disk radius we find (see M15, and references therein).

For the most self-consistent description of the reflection and relativistic blurring we use the \texttt{relxilllp} model, i.e., assuming a lamppost geometry for the corona. While this is a simplified geometry in which the corona is assumed to be a point source  on the spin axis at a given height $H$ above the BH \citep[see, e.g.,][]{dauser13a}, it is the only geometry where the reflection fraction can be calculated self-consistently based on ray-tracing calculations.\footnote{In principle the reflection fraction can be calculated in this way for any geometry \citep[see, e.g.,][]{wilkins12a}, but such calculations are too computationally intensive to be performed while fitting astrophysical data}

This model  also gives an acceptable fit with $\redchi=1.07~(\chi^2=1012)$ for 943 d.o.f.; see Table~\ref{tab:bestfit15}. Compared to the previous model, the reflection fraction is now expressed in terms of coronal height.
We obtain a lower limit for the inner radius $\rin > 35\,\rg$, while the coronal height $H$ is completely unconstrained over the allowed range of 3--100\,$\rg$ (where the lower limit is set by the ISCO for a BH with spin $a=0.8$ and the upper limit is determined to be at a height where changes in $H$ only influence the model marginally). 

\begin{figure}
\includegraphics[width=0.95\columnwidth]{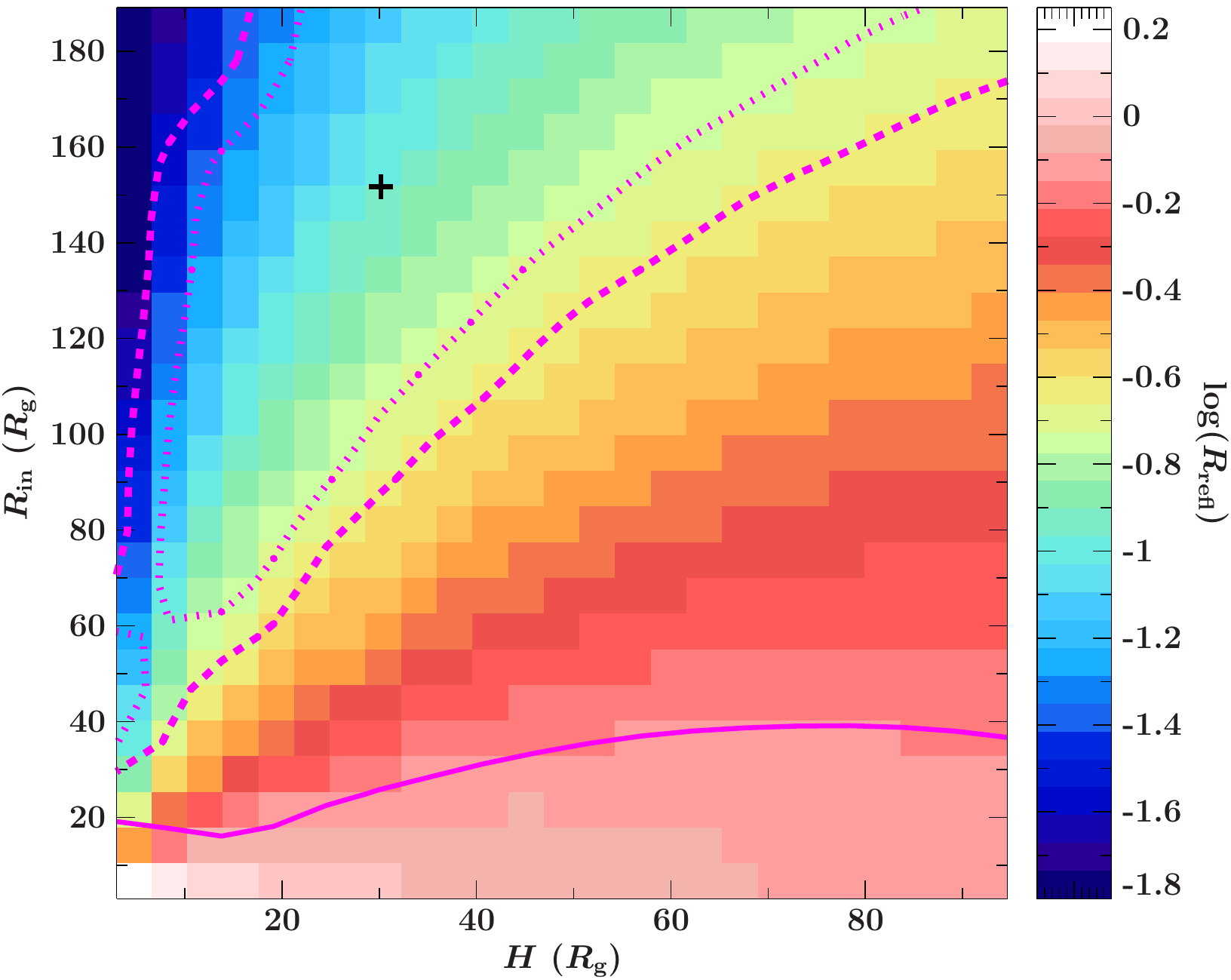}
\caption{Confidence contours of $\chi^2$  for the self-consistent \texttt{relxilllp} model as a function of coronal height $H$ and inner radius $\rin$.  The lines indicate the 1$\sigma$ (dotted), 90\% (dashed) and 99\% (solid) confidence levels for two parameters of interest. The 99\% level only provides a lower limit to the inner radius. The cross marks the best-fit value. The color-coded map in the background shows the corresponding reflection fraction according to the scale on the right.
}
\label{fig:cm_h2rin}
\end{figure}

As both $H$ and \rin are directly related to the reflection fraction, and the reflection fraction is relatively well-constrained, as shown in the \texttt{relxill} model, we expect a strong degeneracy between these parameters. We therefore calculate  a confidence contour between them, shown in Figure~\ref{fig:cm_h2rin}. While this confirms the degeneracy between these two parameters, an inner radius $<17.5\,\rg$ is ruled out at the 99\% confidence level for all values of $H$.

As the reflection fraction is taken into account self-consistently in this model, we can calculate it based on the values for $H$ and $\rin$ (and $a$ and $r_\text{out}$ which have been held fixed). Similar values for the reflection can be achieved over a wide range of values for $H$ and $\rin$, as shown by the color-coded map in the background of Figure~\ref{fig:cm_h2rin}. The confidence contours follow areas of constant reflection fraction closely.

\section{Discussion and conclusion}
\label{sec:disc}
We have presented a spectral analysis of \xmm and \nustar observations of \grs during a very faint hard state. The luminosity between 1--80\,keV was about $3\times10^{35}$\,erg\,s$^{-1}$, i.e., only about 0.02\% of the Eddington luminosity for a prototypical 10\,\msun BH at a distance of 8.5\,kpc.
The \xmm and \nustar spectra agree very well and provide, despite the low source flux, a high-quality spectrum between 0.5--60\,keV. While the reflection features are weak, they are still  detected at $>98\%$ confidence in our data.

The spectrum is very hard with a photon index around 1.4 and a folding energy at $\sim$60\,keV. 
It is somewhat surprising to find such a hard spectrum at the very low Eddington luminosity observed. Typically the  photon index decreases with decreasing flux only down to a transitional luminosity of $\sim$1\%\,\ledd, after which the photon index begins to increase again with lower luminosities \citep[see, e.g.,][]{tomsick01a,wu08a, yang15a}. During quiescence the photon index has been seen to increase to $\Gamma\geq2$ \citep{corbel06a, plotkin13a}.
The lowest photon-indices at the transitional luminosity are typically $\sim1.5$ \citep{kalemci13a, wu08a}. 

We observe a harder photon index at roughly two orders of magnitude below the typically expected transition luminosity. Our inferred Eddington luminosity depends on the  assumption of mass and distance, but even with their large uncertainties, it is difficult to increase the luminosity by two orders of magnitude. 
In any case, the measured hard photon index is at the lower end of known indices and comparable to the hardest spectrum found by 
\citet{belloni02a} for XTE~J1550$-$564. This may indicate that thermal Comptonization in an optically thin plasma is still the dominating effect in \grs, even though a strong radio jet is present (e.g., Loh et al., in prep.), as a jet-dominated synchrotron spectrum would result in a softer photon index \citep{esin97a, yang15a}.

A faint hard state of the prototypical transient BH binary \gx was presented by \citet{gx339}, at an estimated luminosity of 0.94\%\,\ledd.  We found that the spectrum incident on the reflector was harder than the observed continuum, with a best-fit photon index of $\Gamma=1.31^{+0.01}_{-0.31}$. This is similar to the values we measure for \grs. \citet{gx339} argue that the inner parts of the corona, which are preferentially intercepted and reprocessed by the accretion disk, might be hotter than parts farther away from the BH, which are more likely to be visible by a distant observer. If in \grs the accretion disk is truncated or its  inner parts are optically thin, we would have a direct line of sight towards the hot inner parts of the corona, explaining the observed hard power law. 

In \grs we find a relatively low folding energy of $\sim$60\,keV. In the \texttt{nthcomp} Comptonization model we find a corresponding low electron temperature around 16\,keV (resulting in a high optical depth of $\tau > 3$, \citealt{sunyaev80a}). Such a  cool corona is unusual at these low luminosities \citep{tomsick01a, miyakawa08a, gx339}. However, there are a few examples of other BH systems that have shown a low cutoff energy together with a hard photon index \citep[e.g., GRO~J1655$-$40,][]{kalemci16a}. We note that M15 also found a relatively low cutoff energy of 28\,keV, cooler than in our observation. It is therefore possible that \grs has a generally cooler corona than comparable BH binaries.

We applied two relativistic reflection models to the \grs data, with different assumptions: either assuming a constant emissivity index of $q=3$ or a self-consistent emissivity and reflection fraction in the lamppost geometry. In both cases we find a significantly truncated accretion disk at the 90\% confidence limit at $\rin > 15\,\rg$ and $>35 \rg$, respectively. In the self-consistent lamppost model, we can even rule out an accretion disk with an inner radius $\lesssim 20\,\rg$ at the 99\% level. 
However, all  these values are strongly dependent on our assumptions. In the following we will discuss three assumptions influencing  the systematic uncertainties.

\textbf{The coronal and disk geometry}: While the lamppost  geometry is likely a significant simplification of the real geometry (e.g., by assuming a point-like corona), there are strong indications that the X-ray corona is compact, at least at luminosities $L\gtrsim1\%\ledd$ \citep[e.g.,][]{reis13a}. Furthermore, when describing the emissivity with a broken power law, values resembling the lamppost geometry of a corona close to the black hole, i.e., a very steep inner index and a much flatter outer index, are often found \citep[e.g.,][M15]{wilkins12a}. However, the coronal structure in the very low hard state, as observed here, is much less certain, and the applicability of a lamppost corona is unclear. For example, if most parts of the inner accretion disk are replaced by an ADAF, the ADAF itself could act as the Compton upscattering hot electron gas. In this case the inner accretion disk would naturally be truncated as well.

We note that the non-relativistic \texttt{xillver} model provides a good fit to the data and that the relativistic models are consistent with a neutral ionization parameter. This could indicate that the reflection occurs very far away from the BH, maybe in neutral material independent of the accretion disk or possibly on the companion's  surface. This would be possible for strongly beamed and misaligned coronal emission and is also consistent with a strongly truncated accretion disk.

It is possible  that the  corona is outflowing and thereby beaming most of its radiation away from the accretion disk. In this case, we would observe a low reflection fraction despite a non-truncated accretion disk \citep{beloborodov99a}. This model is particularly relevant if the corona is associated with the base of a relativistic jet, which is known to be present due to the strong flux in the radio (Loh et al., in prep.). However, the data quality does not allow us to constrain such an outflow and we can therefore not quantitatively assess this possibility.

\textbf{Inclination}: Here we assume an inclination of $32.5^\circ$, as found by M15 for the lamppost geometry. In the model preferred by M15, with an emissivity described by a broken power law, they find   $43.2^\circ$ instead. When using  this higher inclination, we find a truncated accretion disk at $>28\,\rg$ at the 90\% level, and we can no longer  constrain the radius at the 99\% level, even with the self-consistent lamppost model (i.e., all inner radii between 3--200\,$\rg$ are allowed at the 99\% level). It is possible that the inclination of the accretion disk changed between the two observations, e.g., due to a warped disk \citep[e.g.,][]{tomsick14a}, so that a large range of values is possible. Our data do not allow us to constrain the disk inclination independently.  

\textbf{Outer radius}: as we find that our data are consistent with large values of the inner truncation radius ($\rin \geq 200\,\rg$), we investigate if the choice of the outer accretion disk radius influences the constraints. As the reflection fraction is calculated self-consistently from the size of the accretion disk in the \texttt{relxilllp} model, a change in outer radius will influence the inferred reflection fraction. The typical assumption in most relativistic reflection models is an outer radius of 400\,$\rg$, which is justified for steep emissivity indices. To confirm that this choice does not influence our measurement, we stepped the outer radius from 400\,$\rg$ to 1000\,$\rg$ (the upper limit of the \texttt{relxilllp} model) and find consistent values of $\rin\approx$20\,$\rg$ at the 99\% limit. 

Another important parameter for relativistic reflection models is the BH spin, $a$, which we held fix at 0.8 as found by M15. While this value is not well constrained, changes of the spin do not influence the spectral fits in our case, given the large inner radius we find. Even for a non-spinning BH our lower limits are far outside the ISCO, which would be at 6\,$\rg$. The exact value of the spin parameters therefore does not change our conclusions.

In conclusion we have shown that the combination of \xmm and \nustar allows us to get a more detailed look at BH accretion at  lower Eddington luminosities than ever before. We can constrain the underlying continuum very well and find strong indications that the accretion disk is truncated at a minimum of $15\,\rg$, i.e., $\sim 5\,\risco$ for a BH with spin $a=0.8$. However, even with these data, a unique determination of the geometry of the corona and the accretion disk in this state cannot be found due to the lack of photons as well as strong degeneracies in the models. 

\acknowledgments
We thank the referee for their helpful comments.
We would like to thank the schedulers and SOC of \xmm and \nustar for making these observations possible.
Based on observations obtained with \xmm, an ESA science mission with instruments and contributions directly funded by ESA Member States and NASA.
This work is based upon work supported by NASA under award  No. NNX16AH17G.
JAT acknowledges partial support from NASA under {\em Swift} 
Guest Observer grants NNX15AB81G and NNX15AR52G.
EK acknowledges support of TUBITAK Project No 115F488.
SC and AL acknowledge funding support from the French Research National Agency: CHAOS project ANR-12-BS05-0009 and the UnivEarthS Labex program of Sorbonne Paris Cit\'e (ANR-10-LABX-0023 and ANR-11-IDEX-0005-02).
JCAM-J is the recipient of an Australian Research Council Future Fellowship (FT140101082).
This work was supported under NASA Contract No. NNG08FD60C, and
made use of data from the {\it NuSTAR} mission, a project led by
the California Institute of Technology, managed by the Jet Propulsion
Laboratory, and funded by the National Aeronautics and Space
Administration. We thank the {\it NuSTAR} Operations, Software and
Calibration teams for support with the execution and analysis of these observations. This research has made use of the {\it NuSTAR}
Data Analysis Software (NuSTARDAS) jointly developed by the ASI
Science Data Center (ASDC, Italy) and the California Institute of
Technology (USA). 
We would like to thank John E. Davis for the \texttt{slxfig} module, which was used to produce all figures in this work.
This research has made use of MAXI data provided by RIKEN, JAXA and the MAXI team.
The \swift/BAT transient monitor results were
provided by the \swift/BAT team.
This research has made use of a collection of ISIS functions (ISISscripts) provided by ECAP/Remeis observatory and MIT (\url{http://www.sternwarte.uni-erlangen.de/isis/}).

\textit{Facilities:} \facility{NuSTAR}, \facility{XMM}


\end{document}